\documentclass[fleqn,usenatbib]{mnras}

\usepackage[T1]{fontenc}

\DeclareRobustCommand{\VAN}[3]{#2}
\let\VANthebibliography\thebibliography
\def\thebibliography{\DeclareRobustCommand{\VAN}[3]{##3}\VANthebibliography}

\usepackage{graphicx}	
\usepackage{amsmath}	
\usepackage{amssymb}	
\usepackage{hyperref}

\usepackage{siunitx}
\usepackage{newtxtext,newtxmath}
\usepackage{dsfont}
\usepackage{subcaption}
\usepackage[labelfont=bf, labelsep=period]{caption}
\usepackage{booktabs}
\title[Score-based modelling]{Realistic galaxy image simulation via score-based generative models}

\author[M. J. Smith et al.]
{\parbox{\textwidth}{
Michael J. Smith$^{1, 2}$\thanks{mike@mjjsmith.com},
James E. Geach$^{1, 2}$,
Ryan A. Jackson$^{3}$,
Nikhil Arora$^{4}$,
Connor Stone$^{4}$ \\ and
St\'ephane Courteau${^4}$}\\\\
$^{1}$Centre of Data Innovation Research, Department of Physics, Astronomy, and Mathematics, University of Hertfordshire, Hatfield, AL10 9AB\\
$^{2}$Centre of Astrophysics Research, Department of Physics, Astronomy, and Mathematics, University of Hertfordshire, Hatfield, AL10 9AB\\
$^{3}$Department of Astronomy and Yonsei University Observatory, Yonsei University, Seoul 03722, Republic of Korea\\
$^{4}$Department of Physics, Engineering Physics, and Astronomy, Queen’s University, Kingston, ON K7L 3N6, Canada
}

\pubyear{2022}

\begin{document}

\label{firstpage}
\pagerange{\pageref{firstpage}--\pageref{lastpage}}
\maketitle

\begin{abstract}
    We show that a Denoising Diffusion Probabilistic Model (DDPM), a class of
    score-based generative model, can be used to produce realistic mock
    images that mimic observations of galaxies. Our method is tested with Dark
    Energy Spectroscopic Instrument (DESI) {\it grz} imaging of galaxies from the
    Photometry and Rotation curve OBservations from Extragalactic Surveys
    (PROBES) sample and galaxies selected from the Sloan Digital Sky Survey.
    Subjectively, the generated galaxies are highly realistic when compared
    with samples from the real dataset. We quantify the similarity by
    borrowing from the deep generative learning literature, using the
    `Fr\'{e}chet Inception Distance' to test for subjective and morphological
    similarity. We also introduce the `Synthetic Galaxy Distance' metric
    to compare the emergent physical properties (such as total magnitude,
    colour and half light radius) of a ground truth parent and synthesised
    child dataset. We argue that the DDPM approach produces sharper and more
    realistic images than other generative methods such as adversarial networks
    (with the downside of more costly inference), and could be used to produce
    large samples of synthetic observations tailored to a specific imaging
    survey. We demonstrate two potential uses of the DDPM: (1) accurate
    in-painting of occluded data, such as satellite trails, and (2) domain
    transfer, where new input images can be processed to mimic the properties
    of the DDPM training set.  Here we `DESI-fy' cartoon images as a proof of
    concept for domain transfer.  Finally, we suggest potential applications
    for score-based approaches that could motivate further research on this
    topic within the astronomical community.
\end{abstract}




\section{Introduction}

Synthetic data will play a pivotal role as we journey further into astronomy's
epoch of big data, especially for large extragalactic surveys
\citep{cite_sdss,cite_ska,cite_euclid,cite_lsst}.  It will be required to train
machine learning methods, to interpret observations, and to test theoretical
frameworks. Indeed, one form of synthetic data comes from theoretical models.
For example, in the field of galaxy formation and evolution, simulations using
semi-analytical approaches have been successful in reproducing many of the bulk
observable and emergent properties of galaxies over a significant fraction of
cosmic time
\citep[e.g.][]{cite_somerville1999,cite_cole2000,cite_bower2006,cite_croton2006}.
Semi-analytical models (SAMs) employ approximations derived from more detailed
numerical simulations and empirical calibrations from data to model galaxy
formation and evolution. So it is possible to generate, for example, a `mock'
catalogue of galaxies with predicted optical photometry \citep{cite_lagos2019}.
Hydrodynamical models of galaxy formation track the evolution of baryons and
dark matter within representative volumes
\citep[e.g.][]{cite_dubois2014,cite_vogelsberger2014,cite_eagle,cite_khandai2015,cite_kaviraj2017},
and when pushed to high spatial resolution, can predict galaxy morphologies on
physical scales commensurate with the angular scales achievable with current
observational facilities. When radiative transfer schemes are applied for the
propagation of (for example) starlight through the volume, realistic synthetic
observations can be produced, to be compared with nature
\citep[e.g.][]{cite_camps2016,cite_trayford2017,cite_lovell2021}.

\begin{figure*}
    \includegraphics{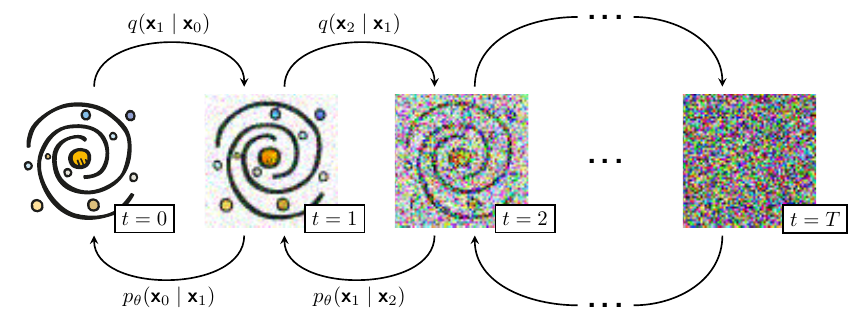}
    \caption{It is easy (and achievable without learnt parameters) to add noise
    to an image, but more difficult to remove it. DDPMs attempt to learn an
    iterative removal process through an appropriate neural network.
    $p_\theta$.}
    \label{fig_ddpm}
\end{figure*}
   
To properly mimic real astronomical data, with all its wonderful
idiosyncrasies, requires a full and detailed understanding of the telescope
response, instrumental properties, and observing conditions, not to mention the
nuances of any data reduction procedure. These non-trivial steps are
typically unique to a given set of observations. There is a short-cut however:
armed with enough examples of observations from a given survey, it should be
possible to derive a data-driven approach to mimic not only the content of
interest -- astronomical signal -- but also the properties of the data themselves.
Deep generative models enable precisely that.

Great attention has been given to applications of deep
generative learning to problems in astronomy lately. 
Generative Adversarial Networks
\citep[GAN;][]{cite_goodfellow2014} have been used for deconvolution
\citep{cite_schawinski2017}, synthetic galaxy generation
\citep{cite_ravanbakhsh2016,cite_fussell2019}, dark matter simulation
\citep{cite_mustafa2017,cite_tamosiunas2020}, and deep field imagery generation
\citep{cite_smith2019}. 
Variational Auto-Encoders
\citep[VAE;][]{cite_kingma2014} have been used to simulate galaxy observations
\citep{cite_ravanbakhsh2016,cite_spindler2021,cite_lanusse2021}, as have
flow-based models \citep{cite_rezende2015,cite_bretonniere2021}. In this paper,
we show that it is possible to simulate realistic galaxy imagery with a Denoising
Diffusion Probablisitic Model (DDPM).  

DDPMs were introduced by \citet{cite_sohldickstein2015} and were first shown to
produce high quality synthetic samples by \citet{cite_ho2020}.
They belong to a family of generative deep learning models that employ
denoising `score matching' via annealed Langevin dynamic sampling
\citep{cite_song2020,cite_ho2020,cite_ajm2020,cite_song2021}.  This family of
score-based generative models (SBGMs) can generate imagery of a quality and
diversity surpassing state of the art GAN models, a startling result
considering the historic disparity in interest and development between the two
techniques \citep{cite_song2021,cite_nichol2021,cite_dhariwal2021}.  

SBGMs have already been used to super-resolve images
\citep{cite_kadkhodaie2020,cite_saharia2021}, translate between image domains
\citep{cite_sasaki2021}, separate superimposed images \citep{cite_jayaram2020},
and in-paint information \citep{cite_kadkhodaie2020,cite_song2021}. At the time
of writing, there is only one example of score-based modelling in the
astronomy literature \citep{cite_remy2020}. This is despite some obvious uses
in astronomical data pipelines. For example: an implementation like
\citet{cite_sasaki2021} could be used for survey-to-survey photometry
translation similarly to \citet{cite_buncher2021}; the source image separation
model described in \citet{cite_jayaram2020} could be applied as an
astronomical object deblender \citep[for
example:][]{cite_stark2018,cite_reiman2019,cite_arcelin2021}; and information
inpainting could be used to remove nuisance objects from observations
\citep{cite_kadkhodaie2020,cite_song2021}.

This paper is organised as follows. Section~\ref{sec_ddpm} introduces the
DDPM formulation used in this paper. In Section~\ref{sec_application} and
Section~\ref{sec_results}, we show that DDPMs are capable of generating diverse
synthetic galaxy observations that are both statistically and qualitatively
indistinguishable from observations found in the training set. We also
demonstrate that DDPMs can in-paint occluded information in an observation,
such as satellite trails\footnote{A
growing problem due to the rapidly increasing population of satellites,
exacerbated by mega-constellations \citep{cite_kocifaj2021}.}, and show that we
can inject realism into entirely unrealistic cartoon imagery. A discussion of
our results and suggestions for future research are presented in
Section~\ref{sec_discussion}.

\section{Denoising diffusion probabilistic models} \label{sec_ddpm}

Denoising Diffusion Probabilistic Models (DDPMs) define a diffusion process
that projects a complex image domain space onto a simple domain space. In the
original formulation, this diffusion process is fixed to a predefined Markov
chain that adds a small amount of Gaussian noise with each step.
Figure~\ref{fig_ddpm} illustrates that this `simple domain space' can be noise
sampled from a Gaussian distribution: $\mathbf{x}_T \sim
\mathcal{N}(\mathbf{0},\mathds{1})$.

\subsection{Forward process}

We define a Markov chain to slowly add Gaussian noise to our data: 
\begin{equation}
    q(\mathbf{x}_{0 \ldots T}) = q(\mathbf{x}_0)\prod^T_{t=1} q(\mathbf{x}_t \mid \mathbf{x}_{t-1}).
\end{equation}
The amount of noise added per step is controlled with a variance schedule
$\{\beta_t \in (0, 1)\}^T_{t=1}$, such that
\begin{equation}
    q(\mathbf{x}_t \mid \mathbf{x}_{t-1}) = \mathcal{N}(\mathbf{x}_t; \sqrt{1 - \beta_t}\,\mathbf{x}_{t-1},\,\beta_t\mathds{1}).
    \label{eqn_forwardbeta}
\end{equation}
This process is applied iteratively to the input image, $\mathbf{x}_0$. 
If we define the above equation to only depend on $\mathbf{x}_0$, we
can immediately calculate an image representation $\mathbf{x}_t$ for any $t$
\citep{cite_ho2020}. Defining $\alpha_t = 1 - \beta_t$ and
$\bar{\alpha}_t = \prod^t_{i=1} \alpha_i$:
\begin{align}
    \mathbf{x}_t &= \sqrt{\alpha_t}\,\mathbf{x}_{t-1} + \sqrt{1 - \alpha_t}\,\mathbf{z}_{t-1}\notag\\
                 &= \sqrt{\alpha_t \alpha_{t-1}}\,\mathbf{x}_{t-2} + \sqrt{(1 - \alpha_t) + \alpha_{t} (1 - \alpha_{t-1})}\,\bar{\mathbf{z}}_{t-2}\notag\\
                 &= \sqrt{\alpha_t \alpha_{t-1} \alpha_{t-2}}\,\mathbf{x}_{t-3} + \sqrt{(1 - \alpha_t \alpha_{t-1}) + \alpha_{t} \alpha_{t-1} (1 - \alpha_{t-2})}\,\bar{\mathbf{z}}_{t-3}\notag\\
                 &= \ldots\notag\\
                 &= \sqrt{\bar{\alpha_t}} \mathbf{x}_{0} + \sqrt{1 - \bar{\alpha}_{t}}\mathbf{z},
\end{align}
where $\mathbf{z} \sim \mathcal{N}(\mathbf{0}, \mathds{1})$ and
$\bar{\mathbf{z}}$ is a combination of Gaussians. Substituting this
expression into Equation~\ref{eqn_forwardbeta} removes the $\mathbf{x}_{t-1}$
dependency and yields
\begin{equation}
    q(\mathbf{x}_t \mid \mathbf{x}_0) = \mathcal{N}(\mathbf{x}_t; \sqrt{\bar{\alpha}_t}\,\mathbf{x}_0,\,(1-\bar{\alpha}_t)\mathds{1}).
\end{equation}

\begin{figure*}
    \includegraphics[width=\linewidth]{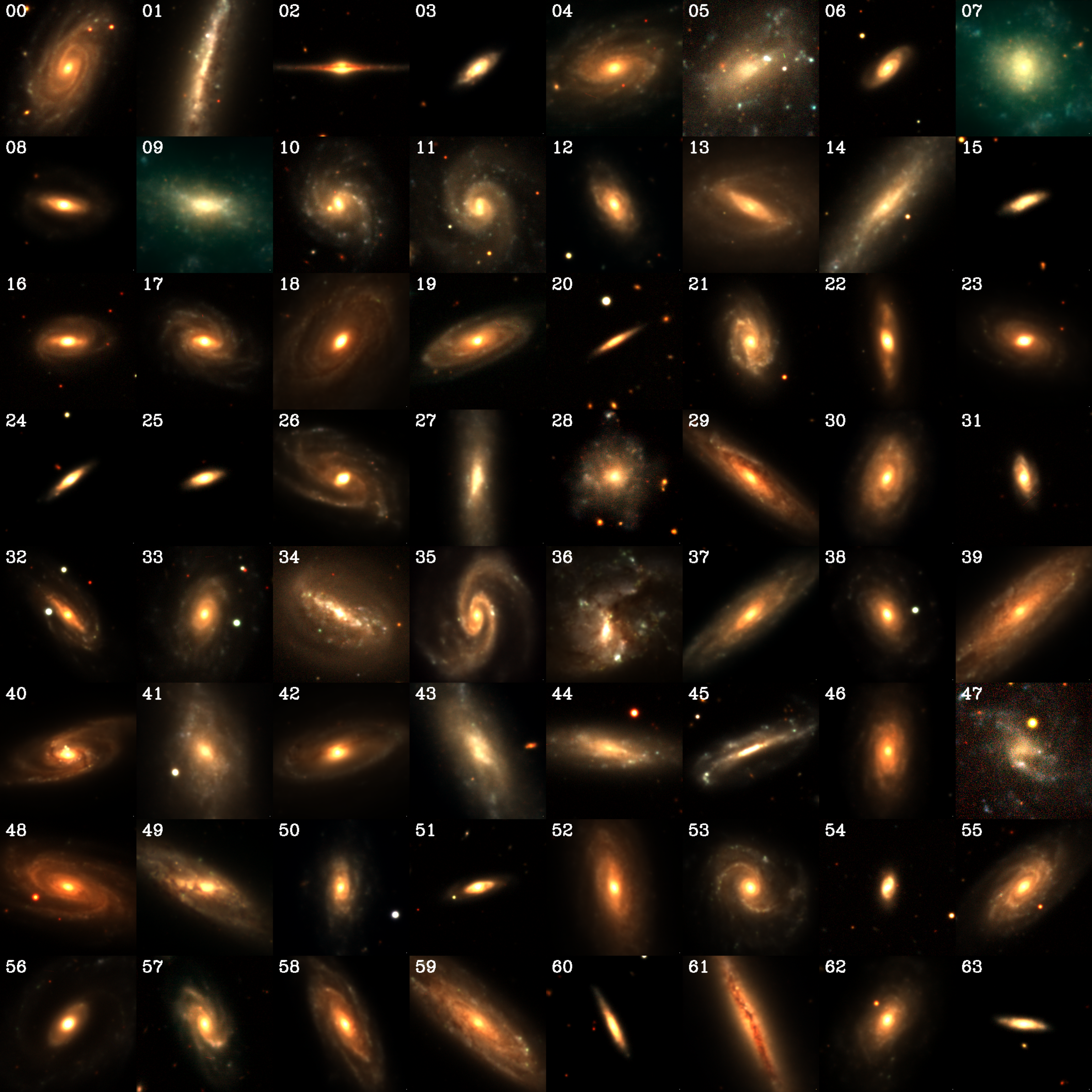}
    \caption{A montage of generated galaxies designed to mimic the PROBES data
    set, interspersed with real examples from the dataset itself. The images
    have been shuffled and the synthetic/real data split is 50/50. All images
    are {\it grz} RGB composites with identical scaling (we have performed a
    99.5\% percentile clip to better show low surface brightness features). A key
    stating which galaxies are real and which are generated is provided at the
    end of this manuscript. More generated galaxies can be found at
    \url{http://mjjsmith.com/thisisnotagalaxy}.}
    \label{FIG_GALAXIES}
\end{figure*}

\subsection{Reverse process}

DDPMs attempt to reverse the forward process by applying a Markov chain with
learnt Gaussian transitions. In our case these transitions are learnt via an
appropriate neural network, $p_\theta$:
\begin{align}
    p_\theta(\mathbf{x}_{0 \ldots T}) &= p(\mathbf{x}_T)\prod^T_{t=1} p_\theta(\mathbf{x}_{t-1} \mid \mathbf{x}_t),\\
    p_\theta(\mathbf{x}_{t - 1} \mid \mathbf{x}_t) &= \mathcal{N}(\mathbf{x}_{t - 1}; \boldsymbol{\mu}_\theta(\mathbf{x}_t, t),\boldsymbol{\Sigma}_\theta(\mathbf{x}_t, t)).
\end{align}
While $\mathbf{\Sigma}_\theta(\mathbf{x}_t, t)$ can be learnt\footnote{See for
example \citet{cite_nichol2021}.}, we follow \citet{cite_ho2020} and fix it to
an iteration-dependent constant $\sigma_t^2 \mathds{1}$, where
\mbox{$\sigma^2_t = 1 - \alpha_t$}.

By recognising that DDPMs are a restricted class of Hierarchical VAE, 
we see that we can train $p_\theta$ by optimising the evidence lower bound
\citep[ELBO, introduced in][]{cite_kingma2014} that can be written as a
summation over the Kullback-Leibler divergences at each iteration
step\footnote{
    See Appendix~B in \citet{cite_sohldickstein2015} and
    Appendix~A in \citet{cite_ho2020} for the full derivation.  
}:
\begin{align}
    \mathcal{L}_{\text{ELBO}} &= \mathbb{E}_q \Bigl[ D_\text{KL}(q(\mathbf{x}_T \mid \mathbf{x}_0) \| p(\mathbf{x}_T))\notag \\
                                &\qquad\quad+ \sum_{t > 1} D_\text{KL}(q(\mathbf{x}_{t - 1} \mid \mathbf{x}_t, \mathbf{x}_0) \| p_\theta(\mathbf{x}_{t - 1} \mid \mathbf{x}_t))\notag \\
                                &\qquad\quad+ \log p_\theta(\mathbf{x}_0 \mid \mathbf{x}_1)\Bigr].
                                \label{eqn_elbo}
\end{align}
In the \citet{cite_ho2020} formulation, the first term in
Equation~\ref{eqn_elbo} is a constant during training and the final term is
modelled as an independent discrete decoder. This leaves the middle summation.
Each term in that summation can be written as
\begin{equation}
    \mathcal{L}(\boldsymbol{\mu}_t, \boldsymbol{\mu}_\theta) = \frac{1}{2 \sigma_t^2} \| \boldsymbol{\mu}_t(\mathbf{x}_t, \mathbf{x}_0) - \boldsymbol{\mu}_\theta(\mathbf{x}_t, t) \|^2,
    \label{eqn_lossmu}
\end{equation}
where $\boldsymbol{\mu}_\theta$ is the neural network's estimation of the forward process
posterior mean $\boldsymbol{\mu}_t$. In practice it would be preferable to predict the noise
addition in each iteration step ($\mathbf{z}_t$), as $\mathbf{z}_t$ has a distribution that by
definition is centred about 0, with a well defined variance.  To this end we
can define $\boldsymbol{\mu}_\theta$ as
\begin{equation}
    \boldsymbol{\mu}_\theta(\mathbf{x}_t, t) = \frac{1}{\sqrt{\alpha_t}} \left(\mathbf{x}_t - \frac{1 - \alpha_t}{\sqrt{1 - \bar{\alpha}_t}} \mathbf{z}_\theta(\mathbf{x}_t, t)\right),
    \label{eqn_mu}
\end{equation}
and by combining Equations~\ref{eqn_lossmu} and \ref{eqn_mu} we get
\begin{align}
    \mathcal{L}(\mathbf{z}_t, \mathbf{z}_\theta) &= 
        \frac{1}{2 \sigma_t^2} \Biggl\| \frac{1}{\sqrt{\alpha_t}} \left(\mathbf{x}_t - \frac{1 - \alpha_t}{\sqrt{1 - \bar{\alpha}_t}} \mathbf{z}_t\right) -\notag \\
         &\qquad\qquad\qquad\qquad\quad \frac{1}{\sqrt{\alpha_t}} \left(\mathbf{x}_t - \frac{1 - \alpha_t}{\sqrt{1 - \bar{\alpha}_t}} \mathbf{z}_\theta(\mathbf{x}_t, t)\right) \Biggr\|^2\notag\\
         &= \frac{(1 - \alpha_t)^2}{2 \sigma_t^2 \alpha_t (1 - \bar{\alpha}_t)} \| \mathbf{z}_t - \mathbf{z}_\theta(\mathbf{x}_t, t) \|^2.
         \label{eqn_complexloss}
\end{align}

\citet{cite_ho2020} empirically found that a simplified version of the loss
described in Equation~\ref{eqn_complexloss} results in better sample quality.
We therefore use a simplified version of Equation~\ref{eqn_complexloss} as our
loss, and optimise to predict the noise required to reverse a forward process
iteration step:
\begin{equation}
    \mathcal{L}(\mathbf{z}_t, \mathbf{z}_\theta) = \| \mathbf{z}_t - \mathbf{z}_\theta (\textbf{x}_t,\,t) \|^2,
    \label{eqn_loss}
\end{equation}
where $\mathbf{x}_t = \sqrt{\bar{\alpha}_t} \mathbf{x}_0 + \sqrt{1 -
\bar{\alpha}_t}\mathbf{z}_t$.

By recognising that $\mathbf{z}_t = \sigma_t^2 \nabla_{\mathbf{x}_t} \log q(\mathbf{x}_t
\mid \mathbf{x}_{t - 1})$, we see that Equation~\ref{eqn_loss} is equivalent to
denoising score matching over $t$ noise levels \citep{cite_vincent2011}.  This
connection establishes a link between DDPMs and other SBGMs \citep[such
as][]{cite_song2019,cite_song2020,cite_ajm2020}.  

Here we use a modified U-Net as $p_\theta$
\citep{cite_ronneberger2015,cite_salimans2017}, and train via the Adam
optimiser \citep{cite_kingma2015}.  The U-Net comprises of three downsample
blocks, a bottleneck block, and three upsample blocks. Each downsample block
comprises of two residual blocks \citep{cite_srivastava2015,cite_he2015}, a
self-attention layer \citep{cite_bahdanau2014,cite_cheng2016}, and a strided
convolution layer. The bottleneck comprises of a self-attention layer
sandwiched by two residual blocks. Each upsample block comprises of two
residual blocks, a self-attention layer, and a transposed convolution layer. As
in a standard U-Net, residual connections link the downsample and upsample
blocks. To provide information about the current iteration step, an embedding
representing the reverse process iteration step is periodically injected into
the U-Net via a summation.  Mish activation is used throughout
\citep{cite_mish}. The full implementation is released under the AGPLv3 licence
and is available at \url{https://github.com/Smith42/astroddpm}.

To run inference for the reverse process, we progressively remove the predicted
noise $\mathbf{z}_\theta$ from our image. The predicted noise is weighted
according to our variance schedule:
\begin{equation}
    \mathbf{x}_{t-1} = \frac{1}{\sqrt{\alpha_t}} \left(\mathbf{x}_t - \frac{1 - \alpha_t}{\sqrt{1 - \bar{\alpha}_t}}\,\mathbf{z}_\theta(\mathbf{x}_t, t)\right) + \boldsymbol{\sigma}_t \mathbf{z}.
\end{equation}

If we take $p(\mathbf{x}_T) \sim \mathcal{N}(\mathbf{x}_T; \mathbf{0},
\mathds{1})$, we can use $p_\theta$ to generate entirely novel data that are
similar, but not identical to, those found in the training set. We can also use
$p_\theta$ to perform image domain translation, and inpainting.
Section~\ref{sec_results} describes these applications in further detail.

\section{Simulating DESI galaxy images} \label{sec_application}

We train our models on minimally processed native resolution (${256
\times 256}$~pixels at 0.262$''$\,pixel$^{-1}$) Dark Energy Spectroscopic Instrument
\citep[DESI;][]{cite_desi} Legacy Survey Data Release 9 galaxy imagery. The
$g$, $r$, and $z$ band images have an average atmospheric seeing of approximately 1$''$.

\begin{figure}
    \centering
    \includegraphics[width=\linewidth]{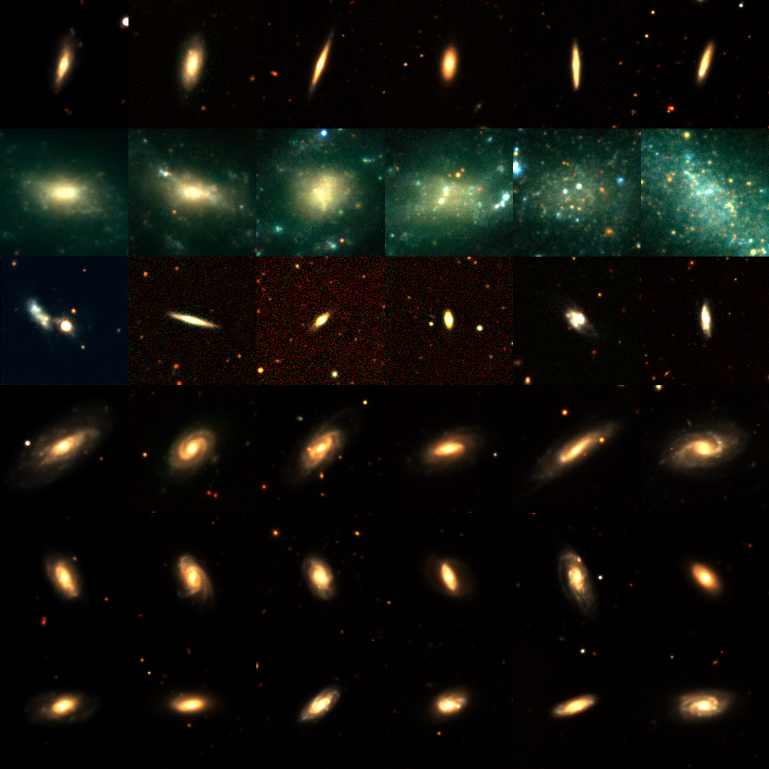}
    \caption{Pixelspace nearest neighbour to generated PROBES galaxies. The leftmost
    column shows a galaxy generated with the model $p_\theta(\mathbf{z})$, the other
    columns show that galaxy's closest training set neighbours in pixelspace.
    Moving along a row takes us further away from
    the simulated galaxy in pixelspace.}
    \label{fig_closest}
\end{figure}

\subsection{Data sample, preparation, and training}

We train two models on two different datasets for 750,000 global steps each
across three NVIDIA Tesla V100 GPUs, corresponding to ${\sim}250$ hours wall
time per model. We fill all available VRAM and set the batch size to 56.
The two datasets are described below.

\begin{enumerate}
    \item We train on the Photometry and Rotation curve OBservations  from
        Extragalactic Surveys (PROBES) galaxy dataset as imaged by the Dark
        Energy Spectroscopic Instrument \citep[DESI;][]{cite_desi} Legacy
        Survey Data Release 9. The PROBES dataset is described in
        \citet{cite_stone2019} and \citet{cite_stone2021a}. It contains 1962
        late-type galaxies with no large neighbours or other obscuring features
        (such as bright stars). Most of the objects are well resolved,
        exhibiting spiral arms, bars, and other features characteristic of
        late-type systems. The model trained on this dataset produces galaxies
        that obviously exhibit internal structure. We refer to this as the
        `PROBES' dataset.

    \item We also train on a dataset of 306\,006 galaxies whose coordinates are
        taken from Sloan Digital Sky Survey \citep[SDSS;][]{cite_sdss} Data
        Release 7 \citep{cite_sdssdr7} and a modified catalogue from
        \cite{cite_wilman2010}. This volume complete sample has an
        $r$-band absolute magnitude limit of $M_r\leq-20$ and a redshift limit
        of $z\leq0.08$. See \citet{cite_arora2019} for details.
        This catalogue covers a wide range of environments from clusters to
        groups and field systems. As in the PROBES dataset, the galaxy images
        are taken from DESI \citep{cite_desi}. We use this dataset and the
        corresponding trained model to compare population level galaxy
        statistics (Sec.~\ref{sec_sgd}). For brevity we refer to this as the
        `SDSS' dataset.
\end{enumerate}

All images are cropped about the target galaxy to a shape of \mbox{$256 \times
256$} pixels. The only destructive pre-processing performed is a upper
and lower percentile clipping, with the percentiles calculated across the
entire dataset. This clipping removes any `hot' or `cold' pixels. To calculate
the upper flux truncation point we evaluate the 99.9th percentile fluxes for
each galaxy across the full dataset. To separate the long tail from
the bulk of the data, we fit a two-cluster {\it k}-means \citep{cite_lloyd1982}.
The two-cluster {\it k}-means returns a boundary at approximately
$5$~analogue-to-digital~unit~(ADU) for the SDSS dataset, and $5.5$~ADU for the
PROBES dataset, and so we set these values as our upper truncation points and
normalisation constants. The lower flux truncation point is set as the minimum
off-source pixel-wise root mean square across the entire dataset. We found this
value to be very close to zero across all bands in both the SDSS and PROBES
datasets, and therefore set the lower flux truncation as $0$~ADU. We apply a
min-max normalisation to the images with the following equation:
\begin{equation}
    \bar{\mathbf{x}} = \frac{2 \cdot \text{max}(0, \text{min}(\mathbf{x}, A))}{A} - 1,
\end{equation}
with $A = 5.0$~ADU being the upper flux truncation for the SDSS dataset, and $A =
5.5$~ADU for the PROBES dataset. We reverse this normalisation when post-processing
inferrals from the model.

\section{Results} \label{sec_results}

Figure~\ref{FIG_GALAXIES} shows a random selection of generated galaxies,
alongside a random selection of real galaxies. The images are shuffled and we
can see that the simulated and real galaxies are subjectively
indistinguishable, at least to the authors (we of course invite the reader to
make their own assessment of fidelity, referring to the answer key given at the
end of this paper). Figure~\ref{fig_closest} presents a random selection of
generated galaxies' nearest neighbours in pixelspace. Since the pixelspace
search does not return identical galaxies, we conclude that the DDPM is not
simply regurgitating imagery, and is indeed generating novel data.  We found a
systematic offset in the simulated pixel fluxes and corrected for it in
post-processing.  To estimate the offset, we calculated the median pixel value
in each of the 10,000 mock and 10,000 real galaxy observations. Each set is
sorted and the medians paired according to their place in the sorted sets.
Finally we fit a linear function to the resulting 10,000 median flux pairs. The
gradient of the fit was used as a scaling factor for the simulated galaxy
images.  We found the multiplier to be $1.18$ in $g$, $1.16$ in $r$, and $1.23$
in $z$ for our DESI observations.  Unfortunately, the exact cause of these
offsets could not be determined.  We propose that this discrepancy is due to a
fundamental property of the neural network and its interaction with sparse
imagery such as our galaxy images.

\subsection{Quantifying similarity} \label{sec_sgd}

\begin{figure}
    \centering
    \includegraphics[width=\linewidth]{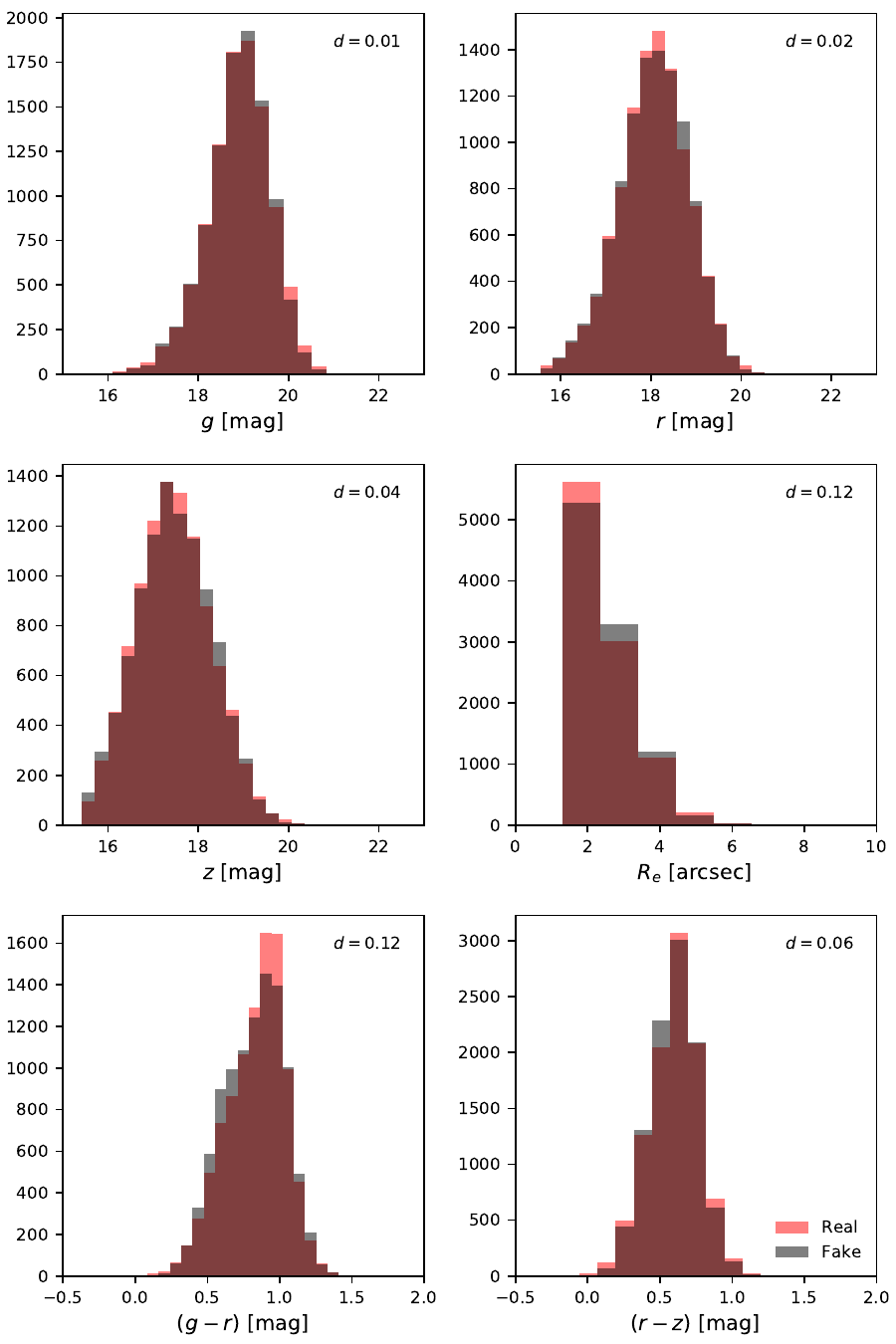}
    \caption{Histogram comparison between galaxies generated by the SDSS DDPM,
    and galaxies contained in the SDSS training set. The half
    light radius histogram follows a lognormal distribution, as do the
    magnitude and colour histograms in flux space. Therefore, we can calculate
    Cohen's $d$ effect size for each histogram pair. As a rule of thumb, if $d
    \le 0.2$ the effect size is considered `small' and a sign of negligible
    difference \citep{cite_cohen1988}.}
    \label{fig_hists_qqs}
\end{figure}

To quantify the similarity of the visual and morphological characteristics of
our galaxies, we borrow from the deep generative learning literature and
calculate the Fr\'echet Inception Distance
\citep[FID;][]{cite_heusel2017,cite_seitzer2020}. The FID is the distance
\citep{cite_dowson1982} between Gaussians fitted to two Inception-v3
\citep{cite_szegedy2016} penultimate layer feature representations. The
penultimate layer nodes are deep in the network and mimic a human's perception
when viewing images. Therefore, if the Gaussians are similar (and the
corresponding FID is small), the images will be visually similar too.

We run FID on 10,000 random samples and present the results in
Table~\ref{tab_results}.  While we cannot yet contextualise our
FID within the literature, we provide the value for future comparison.  
Figure~\ref{FIG_GALAXIES} is presented for a basic visual and morphological
comparison; we cannot discern between the synthesised and real galaxies, which
suggests that the visual and morphological characteristics of our datasets are
well replicated.

To demonstrate that we capture emergent, measurable properties of the galaxies,
we directly compare size and flux distributions. Fluxes are measured via a
summation within a fixed aperture with a diameter of 12~pixels ($\sim$3$''$),
and we use the half light radius as a simple measure of size. To summarise the
distance between the `ground truth' photometry training set properties and the
properties of the simulated set we use the Wasserstein-1 distance\footnote{
    Since we are dealing with large datasets a Kolmogorov-Smirnov
    (KS) test is not appropriate as it becomes overpowered with a very large sample size.  We
    instead use the related Wasserstein-1 distance to provide an absolute value
    that represents the difference between our distribution pairs, and also
    calculate Cohen's $d$ effect size as a direct intuitive substitution for
    the $p$-values that would otherwise result from KS tests
    (Figure~\ref{fig_hists_qqs}).
}:
\begin{equation}
    W(u, v) = \int_{-\infty}^\infty \left\vert U - V \right\vert
    \label{eqn_wasserstein}
\end{equation}
where $U$ and $V$ are the respective cumulative distribution functions of $u$ and $v$.
 
Following Equation~\ref{eqn_wasserstein}, we propose a `synthetic galaxy
distance' metric that captures the difference between emergent properties of a
synthetic and reference galaxy photometry dataset:
\begin{align}
    \text{SGD} &= \sum_i W(u_i, v_i),\notag\\
               &= W(R_e^u, R_e^v) +
                  W(g^u, g^v) +
                  W(r^u, r^v) +
                  W(z^u, z^v)\,+ \notag\\
                  &\hspace{5em}W((g - r)^u, (g - r)^v) +
                  W((r -z)^u, (r - z)^v),
    \label{eqn_sgd}
\end{align}
where $R_e$ is the half light radius, and $g$, $r$, and $z$ are aperture
magnitudes in specific bands and $u$ and $v$ denote different datasets. The SGD
returns a single number, where a lower value denotes a closer match between $u$
and $v$.  When combined with the FID for visual and morphological similarity,
a good overview of the similarity between two large galaxy photometry datasets
is obtained. Figure~\ref{fig_hists_qqs} shows the results for the individual
tests and the SGD summary is in Table~\ref{tab_results}.  We run SGD on 10,000
random samples.

\begin{table}
    \centering
    \caption{Wasserstein-1 distance between emergent property distributions.
    $p_\theta(\mathbf{z})$ is the DDPM described in this paper. `SDSS' is
    a comparison between two different randomly selected sets of 10,000
    galaxies from the training set. We provide the `SDSS' Wasserstein-1 distances
    as a baseline `perfect' inference.}
    \setlength\tabcolsep{3pt}
    \begin{tabular}{l c c c c c c c c}
        \toprule
        & $W_{g}$ & $W_{r}$ & $W_{z}$ & $W_{R_e}$ & $W_{(g - r)}$ & $W_{(r - z)}$ & SGD & FID \\
        \cmidrule(r){1-7} \cmidrule(l){8-9}
        $p_\theta(\mathbf{z})$ & 0.013 & 0.012 & 0.023 & 0.055 & 0.015 & 0.010 & 0.127 & 19 \\
        SDSS & 0.008 & 0.010 & 0.014 & 0.018 & 0.006 & 0.004 & 0.060 & 0.95 \\
        \bottomrule
    \end{tabular}
    \label{tab_results}
\end{table}

\begin{figure}
    \includegraphics[width=\linewidth,angle=0]{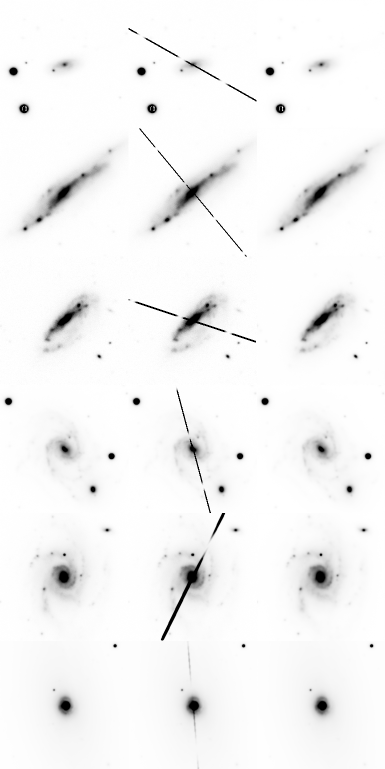}
    \caption{Inpainting of galaxy observations defaced by satellite trails. The
    first column shows the original ({\it r}-band)  PROBES galaxy,
    $\mathbf{x}$. The second column shows the defaced galaxy,
    $\mathbf{\bar{x}}$. The third column shows a random guided draw from the
    model $p_\theta(q(\mathbf{\bar{x}}; T=950))$.}
    \label{fig_sats}
\end{figure}

\begin{figure}
    \centering
    \includegraphics[width=\linewidth]{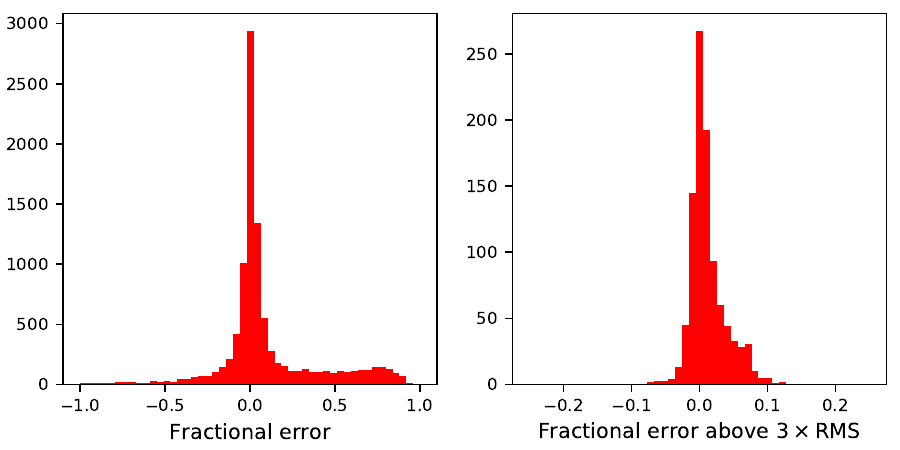}
    \caption{A pixelwise comparison between the ground truth images' occluded
    fluxes, and the recovered images' in Figure~\ref{fig_sats}, expressed as a
    fractional error. The left panel shows the residual for all pixels, and the
    right panel restricts the analysis to pixels exceeding 3 times the
    background r.m.s. ($0.654$~ADU).  We find good agreement
    between the predicted pixel fluxes and the ground truth fluxes, with
    virtually all significant pixels within 10\% of their true values.}
    \label{fig_sats2}
\end{figure}

While we cannot yet contextualise our SGD within the literature, we provide the
value for future comparison. We also present Figure~\ref{fig_hists_qqs} to
otherwise show that our model captures physical properties of the galaxies. We
calculate Cohen's $d$ effect size for each histogram pair, and in all cases
find $d \le 0.2$ indicating a `small' or negligible effect
\citep{cite_cohen1988}. 
Cohen's $d$ effect size is defined here as
\begin{equation}
    d = \left|\sqrt{\frac{2}{\sigma_u^2 - \sigma_v^2}}\left(\mu_u - \mu_v\right)\right|,
\end{equation}
where $\mu$ is the mean of the fitted distribution, and $\sigma$ is the standard
deviation. The subscripts $u$ and $v$ denote different datasets.

\subsection{Satellite trail removal via guided diffusion}

The DDPM can be used to remove simulated galaxy satellite trails from images.
We simulate satellite trails by superimposing a bright linear strip onto a real
image. The strips have a random direction, brightness, width, and periodicity.
In this demonstration we present monochrome {\it r}-band images from the PROBES
dataset.

To perform guided diffusion, we run the reverse process on the occluded part of
the galaxy, in this case the satellite trails. The other image pixels are drawn
directly from the forward process, and so are not updated. The occluded pixels 
are updated with guidance information from the surrounding pixels. 
As Figures~\ref{fig_sats} and \ref{fig_sats2} show, this process
retrieves excellent representations of the original galaxies, essentially
in-painting the missing data with high accuracy. Figure~\ref{fig_sats2} shows
that significant ($>$3$\sigma$) pixels that were occluded by a trail have
recovered fluxes within 10\% of their true values. Since satellite trails are
not present in the training set, a guided draw from the learnt model is
effective at `interpolating' the occluded pixels. A similar approach would work
for other unwanted artefacts, such as glints and ghosts, provided they do not
appear frequently in the training set.

\subsection{Domain transfer} \label{sec_domaintransfer}

\begin{figure}
    \centering
    \includegraphics[width=\linewidth]{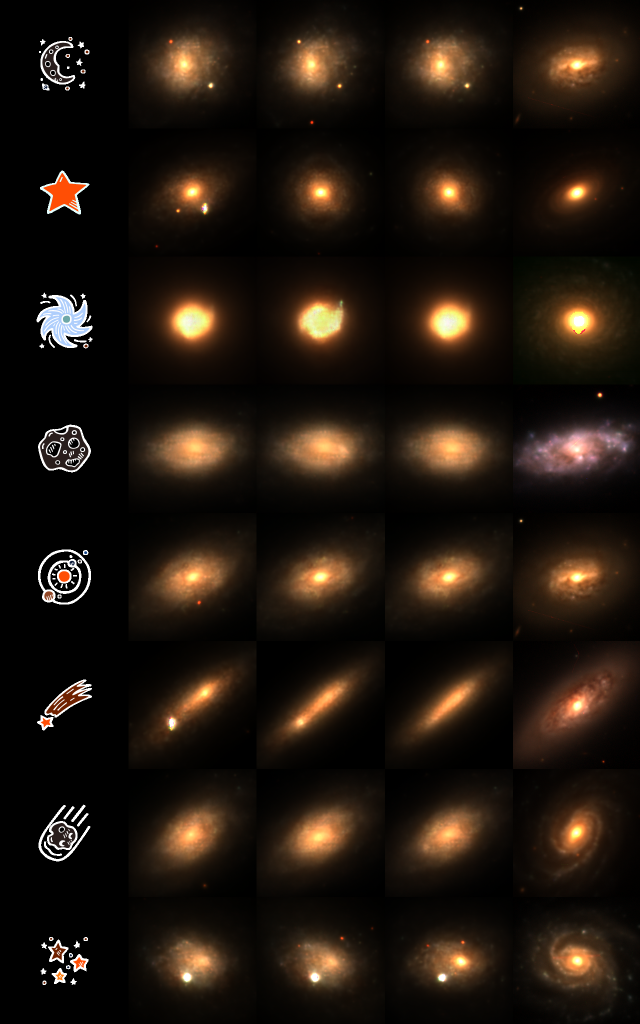}
    \caption{Cartoon images are made to look like DESI observations via
    the model $p_\theta(q(\mathbf{x}; T = 600))$. 
    The first column shows the input image, the middle columns show
    random draws from the PROBES model, and the final column shows
    the pixelspace nearest neighbour to the generated images.}
    \label{fig_transfer}
\end{figure}

The DDPM can also be used to make another input image resemble a DESI Legacy
Survey observation. To perform this domain transfer, we first run the forward
process for $T$ iterations. We then take the noisy image, and run the reverse
process.  This results in a DESI Legacy Survey-like observation that shares
high level features with the input image. Figure~\ref{fig_transfer}
demonstrates this technique on cartoons, setting $T = 600$.  If $T$ is set at a
high value, the DDPM produces an image that more closely resembles one that
might be found in the training set. However, fine detail in the conditioning
image is lost as it is erased by the forward noise addition process.  The
cartoon input is transformed into an image resembling it, but with the
properties of a DESI survey image. Once the cartoon images have been
`DESI-fied', we can search for the nearest neighbour in pixelspace in the real
dataset, and this is shown as a final column in Figure~\ref{fig_transfer}.
Thus, this approach paves the way for pixel-based searching of large survey
imaging databases. For example, one could potentially sketch a particular
morphology or configuration (e.g. an Einstein ring), apply the model tailored
to that survey and then recover the best match in the real data. One could also
apply this technique to inject realism into simulated galaxies, such as those
predicted by hydrodynamical simulations. 

Since the first release of this paper \citet{cite_preechakul2021} and
\citet{cite_saharia2021} have both explored image-to-image translation 
with a score-based model. \citet{cite_preechakul2021} showed that DDPMs 
can produce semantically meaningful embeddings, given an appropriate 
architecture. 
In their paper, they demonstrate that their autoencoding DDPM can 
interpolate along the embedding space and age and de-age images of faces. 
In astronomy, one can imagine using a `survey' embedding to interpolate 
between surveys.

\citet{cite_saharia2021} took a different approach and explicitly trained 
their model to reverse the forward process of an ill-posed inverse problem.   
For an ill-posed inverse problem such as noise addition, one can define 
the forward process in a classical way, and use a DDPM in the inverse 
process to retrieve the uncorrupted image.  
\citet{cite_saharia2021} did this and showed that a DDPM
can colourise greyscale images, and remove JPEG compression artefacts.  

For more difficult problems that do not have a well defined forward 
process, we can use a model similar to that introduced in \citet{cite_sasaki2021} 
to translate between two different image domains. We intend to explore astronomy
related image-to-image translation applications more deeply in follow up
work.

\begin{figure*}     
  \includegraphics[keepaspectratio=true,width=\textwidth]{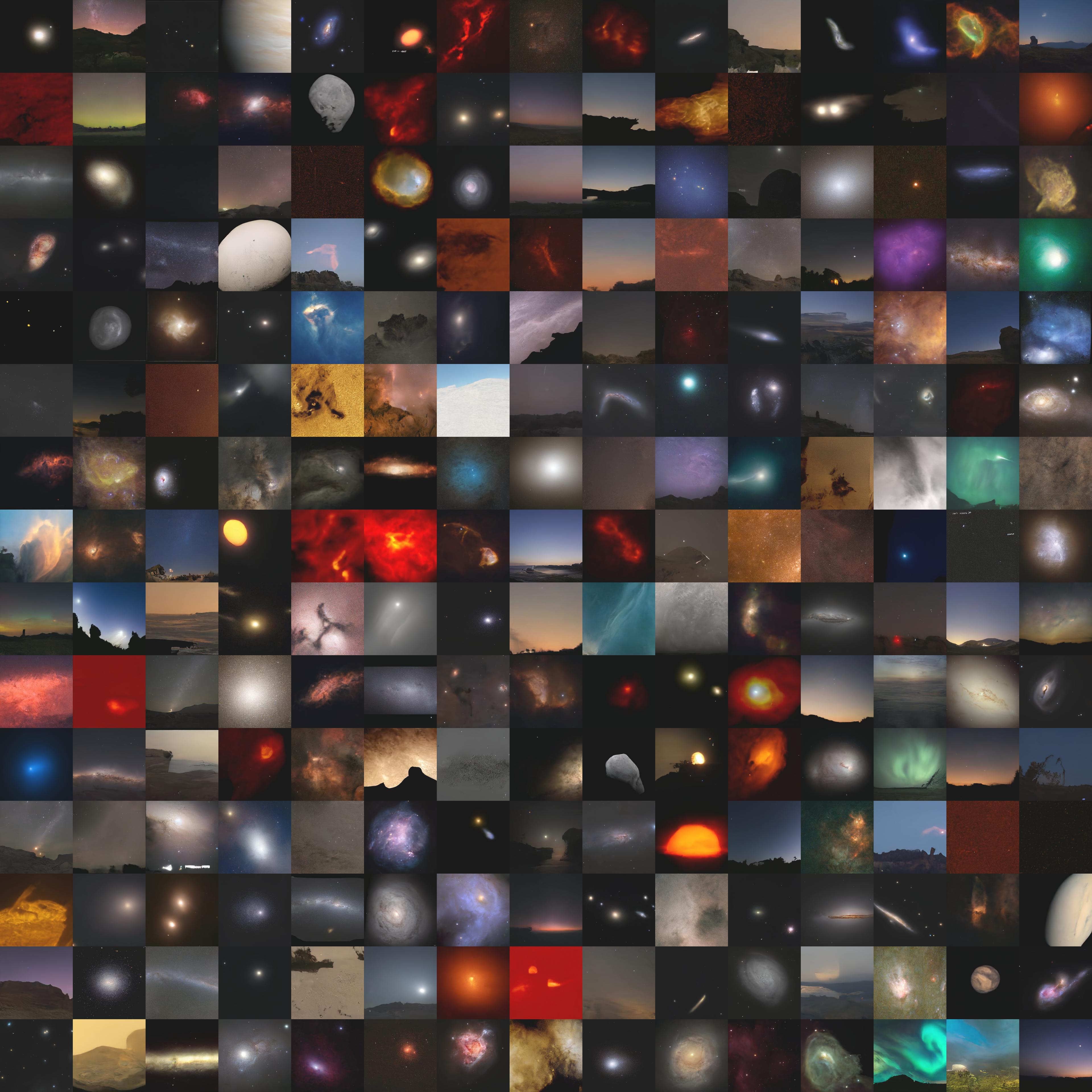}
\caption{A sample of DDPM-generated APOD imagery: AI-APODs. More can be found
    at \url{http://mjjsmith.com/thisisnotanapod} and you can follow a Twitter
    bot \url{https://twitter.com/ThisIsNotAnApod}. A version of this figure has
    been featured on NASA's APOD
    \url{https://apod.nasa.gov/apod/ap211109.html}.}
    \label{fig_apod}
\end{figure*}

\subsection{A fun aside: mock Astronomy Picture of the Day}

As a fun aside, we have trained a DDPM on images from NASA's Astronomy
Picture of the Day (APOD) archive. The dataset comprises 11,428 RGB JPEG
images resized to a $256\times256$ shape. We trained this model for 900,000
global steps on a single V100 GPU. This allows us to generate new APODs that do
not actually exist. Figure~\ref{fig_apod} shows a curated sample of `AI-APODs'
generated from this model. We leave it to the reader to critically assess their
merits, but some common themes are apparent: images resembling nebulae,
galaxies, landscapes, moons, and aurorae are present. Random AI-APODs generated
from the model can be found at \url{http://mjjsmith.com/thisisnotanapod}, and a
Twitter bot will post images at \url{https://twitter.com/ThisIsNotAnApod}.

\section{Conclusions} \label{sec_discussion}

We show that score-based generative modelling is a viable method for synthetic
galaxy image generation, and that this approach preserves emergent properties
such as galaxy size and total flux over different photometric bands, in addition
to producing realistic morphologies. We achieve this fidelity without
explicit encoding of physics or instrumental effects, a great advantage
when simulating physically ill-defined objects. This is a completely
data-driven approach to synthetic data generation.

There are downsides. Naturally, SBGMs require significant computational
resources to train. However, unlike most other generative deep learning methods
SBGMs also require significant resources to infer data. Since they need to
diffuse the data for $T$ cycles\footnote{In this work, $T = 1000$}, it takes
$T$ times longer to produce a batch of synthetic data compared to an equivalent
GAN, VAE, or other single shot generative model. However, there may be routes
to reduce the inference time for SBGMs, with promising results already
\citep{cite_ajm2021,cite_song2021}.

SBGMs have clear astronomical applications, from object deblending
\citep{cite_jayaram2020} to survey-to-survey translation
\citep{cite_sasaki2021} to occluded object in-painting
\citep{cite_kadkhodaie2020,cite_song2021} to super-resolving imagery
\citep{cite_saharia2021}. Unlike GANs, SBGMs do not suffer from mode collapse
and are trivial to train.  SBGMs produce imagery that has a
diversity and fidelity that rivals state of the art GAN models
\citep{cite_song2021,cite_nichol2021,cite_dhariwal2021}. Unlike VAEs,
SBGMs do not constrain information to a fixed bottleneck vector and
thus do not suffer from blurring, and instead produce sharp, realistic imagery
\citep{cite_spindler2021}. For all these reasons, we believe that SBGMs 
are ripe for exploitation by the astronomical community, and we hope 
this paper motivates further work in this topic. 

\section*{Data and code availability}
  
The full PyTorch \citep{cite_pytorch} implementation of the model presented
here is available at \url{https://github.com/Smith42/astroddpm}, and the code
to calculate the Synthetic Galaxy Distance (SGD) can be accessed at
\url{https://github.com/Smith42/synthetic-galaxy-distance}. To calculate the
Fr\'echet Inception Distance (FID), we used
\url{https://github.com/mseitzer/pytorch-fid}.  

\section*{Carbon emissions}

The training of deep learning models requires considerable energy, contributing
to carbon emissions. The energy used to train AstroDDPM to completion is
estimated to be \SI{450}{\kilo\watt\hour}, corresponding to
\SI{105}{\kilo\gram.CO2e} according to the Machine Learning Emissions
Calculator described in \citet{cite_lacoste2019}.  To counteract further
emission from redundant retraining, we follow the recommendations of
\citet{cite_strubell2019} and make available the fully trained models, as well
as the code to run them. Also, we will make available trained models for any
improvements that we make to AstroDDPM in the future.

\section*{Answer key for Figure~\ref{FIG_GALAXIES}}

00 Real, 01 Real, 02 Real, 03 Mock, 04 Real, 05 Real, 06 Real, 07 Mock, 08 Mock, 09 Mock, 10 Real, 11 Real, 12 Mock, 13 Mock, 14 Mock, 15 Mock, 16 Real, 17 Real, 18 Mock, 19 Real, 20 Real, 21 Real, 22 Real, 23 Mock, 24 Mock, 25 Mock, 26 Mock, 27 Mock, 28 Mock, 29 Real, 30 Mock, 31 Real, 32 Real, 33 Real, 34 Real, 35 Mock, 36 Real, 37 Mock, 38 Mock, 39 Mock, 40 Mock, 41 Mock, 42 Mock, 43 Mock, 44 Real, 45 Real, 46 Mock, 47 Real, 48 Real, 49 Real, 50 Mock, 51 Real, 52 Mock, 53 Real, 54 Real, 55 Real, 56 Real, 57 Real, 58 Mock, 59 Real, 60 Mock, 61 Mock, 62 Mock, 63 Mock

\section*{Acknowledgements}

JEG is supported by the Royal Society. This research made use of the University
of Hertfordshire's High Performance Computing facility
(\url{http://uhhpc.herts.ac.uk/}).  The galaxy icon in Figure~\ref{fig_ddpm},
and the cartoons in Figure~\ref{fig_transfer} are by Agata Kuczmi\'nska and are
available under the \mbox{CC-BY-4.0} licence at
\url{https://goodstuffnononsense.com/hand-drawn-icons/space-icons/}. We are
grateful to the Natural Sciences and Engineering Research Council of Canada,
the Ontario Government, and Queen’s University for critical support through
various scholarships and grants. This work was supported in part by the Yonsei
University Research Fund (Yonsei Frontier Lab. Young Researcher Supporting
Program) of 2021. RAJ acknowledges support from the Korean National Research
Foundation (NRF-2020R1A2C3003769). We thank the MNRAS reviewer for helpful
comments and suggestions.

The Legacy Surveys consist of three individual and complementary projects: the
Dark Energy Camera Legacy Survey (DECaLS; Proposal ID \#2014B-0404; PIs: David
Schlegel and Arjun Dey), the Beijing-Arizona Sky Survey (BASS; NOAO Prop. ID
\#2015A-0801; PIs: Zhou Xu and Xiaohui Fan), and the Mayall z-band Legacy Survey
(MzLS; Prop. ID \#2016A-0453; PI: Arjun Dey). DECaLS, BASS and MzLS together
include data obtained, respectively, at the Blanco telescope, Cerro Tololo
Inter-American Observatory, NSF’s NOIRLab; the Bok telescope, Steward
Observatory, University of Arizona; and the Mayall telescope, Kitt Peak
National Observatory, NOIRLab. The Legacy Surveys project is honoured to be
permitted to conduct astronomical research on Iolkam Du’ag (Kitt Peak), a
mountain with particular significance to the Tohono O’odham Nation.

NOIRLab is operated by the Association of Universities for Research in
Astronomy (AURA) under a cooperative agreement with the National Science
Foundation.

This project used data obtained with the Dark Energy Camera (DECam), which was
constructed by the Dark Energy Survey (DES) collaboration. Funding for the DES
Projects has been provided by the U.S. Department of Energy, the U.S. National
Science Foundation, the Ministry of Science and Education of Spain, the Science
and Technology Facilities Council of the United Kingdom, the Higher Education
Funding Council for England, the National Center for Supercomputing
Applications at the University of Illinois at Urbana-Champaign, the Kavli
Institute of Cosmological Physics at the University of Chicago, Center for
Cosmology and Astro-Particle Physics at the Ohio State University, the Mitchell
Institute for Fundamental Physics and Astronomy at Texas A\&M University,
Financiadora de Estudos e Projetos, Fundacao Carlos Chagas Filho de Amparo,
Financiadora de Estudos e Projetos, Fundacao Carlos Chagas Filho de Amparo a
Pesquisa do Estado do Rio de Janeiro, Conselho Nacional de Desenvolvimento
Cientifico e Tecnologico and the Ministerio da Ciencia, Tecnologia e Inovacao,
the Deutsche Forschungsgemeinschaft and the Collaborating Institutions in the
Dark Energy Survey. The Collaborating Institutions are Argonne National
Laboratory, the University of California at Santa Cruz, the University of
Cambridge, Centro de Investigaciones Energeticas, Medioambientales y
Tecnologicas-Madrid, the University of Chicago, University College London, the
DES-Brazil Consortium, the University of Edinburgh, the Eidgenossische
Technische Hochschule (ETH) Zurich, Fermi National Accelerator Laboratory, the
University of Illinois at Urbana-Champaign, the Institut de Ciencies de l’Espai
(IEEC/CSIC), the Institut de Fisica d’Altes Energies, Lawrence Berkeley
National Laboratory, the Ludwig Maximilians Universitat Munchen and the
associated Excellence Cluster Universe, the University of Michigan, NSF’s
NOIRLab, the University of Nottingham, the Ohio State University, the
University of Pennsylvania, the University of Portsmouth, SLAC National
Accelerator Laboratory, Stanford University, the University of Sussex, and
Texas A\&M University.

BASS is a key project of the Telescope Access Program (TAP), which has been
funded by the National Astronomical Observatories of China, the Chinese Academy
of Sciences (the Strategic Priority Research Program “The Emergence of
Cosmological Structures” Grant \# XDB09000000), and the Special Fund for
Astronomy from the Ministry of Finance. The BASS is also supported by the
External Cooperation Program of Chinese Academy of Sciences (Grant \#
114A11KYSB20160057), and Chinese National Natural Science Foundation (Grant \#
11433005).

The Legacy Survey team makes use of data products from the Near-Earth Object
Wide-field Infrared Survey Explorer (NEOWISE), which is a project of the Jet
Propulsion Laboratory/California Institute of Technology. NEOWISE is funded by
the National Aeronautics and Space Administration.

The Legacy Surveys imaging of the DESI footprint is supported by the Director,
Office of Science, Office of High Energy Physics of the U.S. Department of
Energy under Contract No. DE-AC02-05CH1123, by the National Energy Research
Scientific Computing Center, a DOE Office of Science User Facility under the
same contract; and by the U.S. National Science Foundation, Division of
Astronomical Sciences under Contract No. AST-0950945 to NOAO.



\bibliographystyle{mnras}
\bibliography{mnras_template} 

\begin{thebibliography}{}
\makeatletter
\relax
\def\mn@urlcharsother{\let\do\@makeother \do\$\do\&\do\#\do\^\do\_\do\%\do\~}
\def\mn@doi{\begingroup\mn@urlcharsother \@ifnextchar [ {\mn@doi@}
  {\mn@doi@[]}}
\def\mn@doi@[#1]#2{\def\@tempa{#1}\ifx\@tempa\@empty \href
  {http://dx.doi.org/#2} {doi:#2}\else \href {http://dx.doi.org/#2} {#1}\fi
  \endgroup}
\def\mn@eprint#1#2{\mn@eprint@#1:#2::\@nil}
\def\mn@eprint@arXiv#1{\href {http://arxiv.org/abs/#1} {{\tt arXiv:#1}}}
\def\mn@eprint@dblp#1{\href {http://dblp.uni-trier.de/rec/bibtex/#1.xml}
  {dblp:#1}}
\def\mn@eprint@#1:#2:#3:#4\@nil{\def\@tempa {#1}\def\@tempb {#2}\def\@tempc
  {#3}\ifx \@tempc \@empty \let \@tempc \@tempb \let \@tempb \@tempa \fi \ifx
  \@tempb \@empty \def\@tempb {arXiv}\fi \@ifundefined
  {mn@eprint@\@tempb}{\@tempb:\@tempc}{\expandafter \expandafter \csname
  mn@eprint@\@tempb\endcsname \expandafter{\@tempc}}}

\bibitem[\protect\citeauthoryear{Abazajian et~al.,}{Abazajian
  et~al.}{2009}]{cite_sdssdr7}
Abazajian K.~N.,  et~al., 2009, \mn@doi [The Astrophysical Journal Supplement
  Series] {10.1088/0067-0049/182/2/543}, 182, 543

\bibitem[\protect\citeauthoryear{{Amiaux} et~al.,}{{Amiaux}
  et~al.}{2012}]{cite_euclid}
{Amiaux} J.,  et~al., 2012, in {Clampin} M.~C.,  {Fazio} G.~G.,  {MacEwen}
  H.~A.,   {Oschmann} Jacobus~M. J.,  eds,  Society of Photo-Optical
  Instrumentation Engineers (SPIE) Conference Series Vol. 8442, Space
  Telescopes and Instrumentation 2012: Optical, Infrared, and Millimeter Wave.
  p. 84420Z (\mn@eprint {arXiv} {1209.2228}), \mn@doi{10.1117/12.926513}

\bibitem[\protect\citeauthoryear{{Arcelin}, {Doux}, {Aubourg}, {Roucelle}  \&
  {LSST Dark Energy Science Collaboration}}{{Arcelin}
  et~al.}{2021}]{cite_arcelin2021}
{Arcelin} B.,  {Doux} C.,  {Aubourg} E.,  {Roucelle} C.,   {LSST Dark Energy
  Science Collaboration} 2021, \mn@doi [\mnras] {10.1093/mnras/staa3062}, \href
  {https://ui.adsabs.harvard.edu/abs/2021MNRAS.500..531A} {500, 531}

\bibitem[\protect\citeauthoryear{{Arora}, {Fossati}, {Fontanot}, {Hirschmann}
  \& {Wilman}}{{Arora} et~al.}{2019}]{cite_arora2019}
{Arora} N.,  {Fossati} M.,  {Fontanot} F.,  {Hirschmann} M.,   {Wilman} D.~J.,
  2019, \mn@doi [\mnras] {10.1093/mnras/stz2266}, \href
  {https://ui.adsabs.harvard.edu/abs/2019MNRAS.489.1606A} {489, 1606}

\bibitem[\protect\citeauthoryear{{Bahdanau}, {Cho}  \& {Bengio}}{{Bahdanau}
  et~al.}{2014}]{cite_bahdanau2014}
{Bahdanau} D.,  {Cho} K.,   {Bengio} Y.,  2014, CoRR, \href
  {https://ui.adsabs.harvard.edu/abs/2014arXiv1409.0473B} {abs/1409.0473}

\bibitem[\protect\citeauthoryear{{Bower}, {Benson}, {Malbon}, {Helly}, {Frenk},
  {Baugh}, {Cole}  \& {Lacey}}{{Bower} et~al.}{2006}]{cite_bower2006}
{Bower} R.~G.,  {Benson} A.~J.,  {Malbon} R.,  {Helly} J.~C.,  {Frenk} C.~S.,
  {Baugh} C.~M.,  {Cole} S.,   {Lacey} C.~G.,  2006, \mn@doi [\mnras]
  {10.1111/j.1365-2966.2006.10519.x}, \href
  {https://ui.adsabs.harvard.edu/abs/2006MNRAS.370..645B} {370, 645}

\bibitem[\protect\citeauthoryear{{Bretonni{\`e}re} et~al.,}{{Bretonni{\`e}re}
  et~al.}{2021}]{cite_bretonniere2021}
{Bretonni{\`e}re} H.,  et~al., 2021, arXiv e-prints, \href
  {https://ui.adsabs.harvard.edu/abs/2021arXiv210512149B} {p. arXiv:2105.12149}

\bibitem[\protect\citeauthoryear{{Buncher}, {Sharma}  \& {Carrasco
  Kind}}{{Buncher} et~al.}{2021}]{cite_buncher2021}
{Buncher} B.,  {Sharma} A.~N.,   {Carrasco Kind} M.,  2021, \mn@doi [\mnras]
  {10.1093/mnras/stab294}, \href
  {https://ui.adsabs.harvard.edu/abs/2021MNRAS.503..777B} {503, 777}

\bibitem[\protect\citeauthoryear{{Camps}, {Trayford}, {Baes}, {Theuns},
  {Schaller}  \& {Schaye}}{{Camps} et~al.}{2016}]{cite_camps2016}
{Camps} P.,  {Trayford} J.~W.,  {Baes} M.,  {Theuns} T.,  {Schaller} M.,
  {Schaye} J.,  2016, \mn@doi [\mnras] {10.1093/mnras/stw1735}, \href
  {https://ui.adsabs.harvard.edu/abs/2016MNRAS.462.1057C} {462, 1057}

\bibitem[\protect\citeauthoryear{Cheng, Dong  \& Lapata}{Cheng
  et~al.}{2016}]{cite_cheng2016}
Cheng J.,  Dong L.,   Lapata M.,  2016, CoRR, abs/1601.06733

\bibitem[\protect\citeauthoryear{Cohen}{Cohen}{1988}]{cite_cohen1988}
Cohen J.,  1988, {Statistical Power Analysis for the Behavioral Sciences}.
Lawrence Erlbaum Associates

\bibitem[\protect\citeauthoryear{{Cole}, {Lacey}, {Baugh}  \& {Frenk}}{{Cole}
  et~al.}{2000}]{cite_cole2000}
{Cole} S.,  {Lacey} C.~G.,  {Baugh} C.~M.,   {Frenk} C.~S.,  2000, \mn@doi
  [\mnras] {10.1046/j.1365-8711.2000.03879.x}, \href
  {https://ui.adsabs.harvard.edu/abs/2000MNRAS.319..168C} {319, 168}

\bibitem[\protect\citeauthoryear{{Croton} et~al.,}{{Croton}
  et~al.}{2006}]{cite_croton2006}
{Croton} D.~J.,  et~al., 2006, \mn@doi [\mnras]
  {10.1111/j.1365-2966.2005.09675.x}, \href
  {https://ui.adsabs.harvard.edu/abs/2006MNRAS.365...11C} {365, 11}

\bibitem[\protect\citeauthoryear{{Dewdney}, {Hall}, {Schilizzi}  \&
  {Lazio}}{{Dewdney} et~al.}{2009}]{cite_ska}
{Dewdney} P.~E.,  {Hall} P.~J.,  {Schilizzi} R.~T.,   {Lazio} T.~J.~L.~W.,
  2009, \mn@doi [IEEE Proceedings] {10.1109/JPROC.2009.2021005}, \href
  {https://ui.adsabs.harvard.edu/abs/2009IEEEP..97.1482D} {97, 1482}

\bibitem[\protect\citeauthoryear{{Dey} et~al.,}{{Dey} et~al.}{2019}]{cite_desi}
{Dey} A.,  et~al., 2019, \mn@doi [\aj] {10.3847/1538-3881/ab089d}, \href
  {https://ui.adsabs.harvard.edu/abs/2019AJ....157..168D} {157, 168}

\bibitem[\protect\citeauthoryear{Dhariwal \& Nichol}{Dhariwal \&
  Nichol}{2021}]{cite_dhariwal2021}
Dhariwal P.,  Nichol A.,  2021, CoRR, abs/2105.05233

\bibitem[\protect\citeauthoryear{Dowson \& Landau}{Dowson \&
  Landau}{1982}]{cite_dowson1982}
Dowson D.,  Landau B.,  1982, \mn@doi [Journal of Multivariate Analysis]
  {https://doi.org/10.1016/0047-259X(82)90077-X}, 12, 450

\bibitem[\protect\citeauthoryear{{Dubois} et~al.,}{{Dubois}
  et~al.}{2014}]{cite_dubois2014}
{Dubois} Y.,  et~al., 2014, \mn@doi [\mnras] {10.1093/mnras/stu1227}, \href
  {http://adsabs.harvard.edu/abs/2014MNRAS.444.1453D} {444, 1453}

\bibitem[\protect\citeauthoryear{{Fussell} \& {Moews}}{{Fussell} \&
  {Moews}}{2019}]{cite_fussell2019}
{Fussell} L.,  {Moews} B.,  2019, \mn@doi [\mnras] {10.1093/mnras/stz602},
  \href {https://ui.adsabs.harvard.edu/abs/2019MNRAS.485.3203F} {485, 3203}

\bibitem[\protect\citeauthoryear{Goodfellow, Pouget-Abadie, Mirza, Xu,
  Warde-Farley, Ozair, Courville  \& Bengio}{Goodfellow
  et~al.}{2014}]{cite_goodfellow2014}
Goodfellow I.,  Pouget-Abadie J.,  Mirza M.,  Xu B.,  Warde-Farley D.,  Ozair
  S.,  Courville A.,   Bengio Y.,  2014, in Ghahramani Z.,  Welling M.,  Cortes
  C.,  Lawrence N.,   Weinberger K.~Q.,  eds, ~ Vol. 27, Advances in Neural
  Information Processing Systems. Curran Associates, Inc., \url
  {https://proceedings.neurips.cc/paper/2014/file/5ca3e9b122f61f8f06494c97b1afccf3-Paper.pdf}

\bibitem[\protect\citeauthoryear{He, Zhang, Ren  \& Sun}{He
  et~al.}{2015}]{cite_he2015}
He K.,  Zhang X.,  Ren S.,   Sun J.,  2015, CoRR, abs/1512.03385

\bibitem[\protect\citeauthoryear{Heusel, Ramsauer, Unterthiner, Nessler  \&
  Hochreiter}{Heusel et~al.}{2017}]{cite_heusel2017}
Heusel M.,  Ramsauer H.,  Unterthiner T.,  Nessler B.,   Hochreiter S.,  2017,
  in Guyon I.,  Luxburg U.~V.,  Bengio S.,  Wallach H.,  Fergus R.,
  Vishwanathan S.,   Garnett R.,  eds, ~ Vol. 30, Advances in Neural
  Information Processing Systems. Curran Associates, Inc., \url
  {https://proceedings.neurips.cc/paper/2017/file/8a1d694707eb0fefe65871369074926d-Paper.pdf}

\bibitem[\protect\citeauthoryear{Ho, Jain  \& Abbeel}{Ho
  et~al.}{2020}]{cite_ho2020}
Ho J.,  Jain A.,   Abbeel P.,  2020, in Larochelle H.,  Ranzato M.,  Hadsell
  R.,  Balcan M.~F.,   Lin H.,  eds, ~ Vol. 33, Advances in Neural Information
  Processing Systems. Curran Associates, Inc., pp 6840--6851

\bibitem[\protect\citeauthoryear{{Ivezi{\'c}} et~al.,}{{Ivezi{\'c}}
  et~al.}{2019}]{cite_lsst}
{Ivezi{\'c}} {\v{Z}}.,  et~al., 2019, \mn@doi [\apj]
  {10.3847/1538-4357/ab042c}, \href
  {https://ui.adsabs.harvard.edu/abs/2019ApJ...873..111I} {873, 111}

\bibitem[\protect\citeauthoryear{Jayaram \& Thickstun}{Jayaram \&
  Thickstun}{2020}]{cite_jayaram2020}
Jayaram V.,  Thickstun J.,  2020, CoRR, abs/2002.07942

\bibitem[\protect\citeauthoryear{Jolicoeur{-}Martineau,
  Pich{\'{e}}{-}Taillefer, des Combes  \& Mitliagkas}{Jolicoeur{-}Martineau
  et~al.}{2020}]{cite_ajm2020}
Jolicoeur{-}Martineau A.,  Pich{\'{e}}{-}Taillefer R.,  des Combes R.~T.,
  Mitliagkas I.,  2020, CoRR, abs/2009.05475

\bibitem[\protect\citeauthoryear{Jolicoeur{-}Martineau, Li,
  Pich{\'{e}}{-}Taillefer, Kachman  \& Mitliagkas}{Jolicoeur{-}Martineau
  et~al.}{2021}]{cite_ajm2021}
Jolicoeur{-}Martineau A.,  Li K.,  Pich{\'{e}}{-}Taillefer R.,  Kachman T.,
  Mitliagkas I.,  2021, CoRR, abs/2105.14080

\bibitem[\protect\citeauthoryear{Kadkhodaie \& Simoncelli}{Kadkhodaie \&
  Simoncelli}{2020}]{cite_kadkhodaie2020}
Kadkhodaie Z.,  Simoncelli E.~P.,  2020, CoRR, abs/2007.13640

\bibitem[\protect\citeauthoryear{{Kaviraj} et~al.,}{{Kaviraj}
  et~al.}{2017}]{cite_kaviraj2017}
{Kaviraj} S.,  et~al., 2017, \mn@doi [\mnras] {10.1093/mnras/stx126}, \href
  {http://adsabs.harvard.edu/abs/2017MNRAS.467.4739K} {467, 4739}

\bibitem[\protect\citeauthoryear{{Khandai}, {Di Matteo}, {Croft}, {Wilkins},
  {Feng}, {Tucker}, {DeGraf}  \& {Liu}}{{Khandai}
  et~al.}{2015}]{cite_khandai2015}
{Khandai} N.,  {Di Matteo} T.,  {Croft} R.,  {Wilkins} S.,  {Feng} Y.,
  {Tucker} E.,  {DeGraf} C.,   {Liu} M.-S.,  2015, \mn@doi [\mnras]
  {10.1093/mnras/stv627}, \href
  {https://ui.adsabs.harvard.edu/abs/2015MNRAS.450.1349K} {450, 1349}

\bibitem[\protect\citeauthoryear{Kingma \& Ba}{Kingma \&
  Ba}{2015}]{cite_kingma2015}
Kingma D.~P.,  Ba J.,  2015, in Bengio Y.,  LeCun Y.,  eds, 3rd International
  Conference on Learning Representations, {ICLR} 2015, San Diego, CA, USA, May
  7-9, 2015, Conference Track Proceedings. \url
  {http://arxiv.org/abs/1412.6980}

\bibitem[\protect\citeauthoryear{Kingma \& Welling}{Kingma \&
  Welling}{2014}]{cite_kingma2014}
Kingma D.~P.,  Welling M.,  2014, in Bengio Y.,  LeCun Y.,  eds, 2nd
  International Conference on Learning Representations, {ICLR} 2014, Banff, AB,
  Canada, April 14-16, 2014, Conference Track Proceedings. \url
  {http://arxiv.org/abs/1312.6114}

\bibitem[\protect\citeauthoryear{{Kocifaj}, {Kundracik}, {Barentine}  \&
  {Bar{\'a}}}{{Kocifaj} et~al.}{2021}]{cite_kocifaj2021}
{Kocifaj} M.,  {Kundracik} F.,  {Barentine} J.~C.,   {Bar{\'a}} S.,  2021,
  \mn@doi [\mnras] {10.1093/mnrasl/slab030}, \href
  {https://ui.adsabs.harvard.edu/abs/2021MNRAS.504L..40K} {504, L40}

\bibitem[\protect\citeauthoryear{Lacoste, Luccioni, Schmidt  \&
  Dandres}{Lacoste et~al.}{2019}]{cite_lacoste2019}
Lacoste A.,  Luccioni A.,  Schmidt V.,   Dandres T.,  2019, CoRR,
  abs/1910.09700

\bibitem[\protect\citeauthoryear{{Lagos} et~al.,}{{Lagos}
  et~al.}{2019}]{cite_lagos2019}
{Lagos} C. d.~P.,  et~al., 2019, \mn@doi [\mnras] {10.1093/mnras/stz2427},
  \href {https://ui.adsabs.harvard.edu/abs/2019MNRAS.489.4196L} {489, 4196}

\bibitem[\protect\citeauthoryear{{Lanusse}, {Mandelbaum}, {Ravanbakhsh}, {Li},
  {Freeman}  \& {P{\'o}czos}}{{Lanusse} et~al.}{2021}]{cite_lanusse2021}
{Lanusse} F.,  {Mandelbaum} R.,  {Ravanbakhsh} S.,  {Li} C.-L.,  {Freeman} P.,
   {P{\'o}czos} B.,  2021, \mn@doi [\mnras] {10.1093/mnras/stab1214}, \href
  {https://ui.adsabs.harvard.edu/abs/2021MNRAS.504.5543L} {504, 5543}

\bibitem[\protect\citeauthoryear{Lloyd}{Lloyd}{1982}]{cite_lloyd1982}
Lloyd S.,  1982, \mn@doi [IEEE Transactions on Information Theory]
  {10.1109/TIT.1982.1056489}, 28, 129

\bibitem[\protect\citeauthoryear{{Lovell}, {Geach}, {Dav{\'e}}, {Narayanan}  \&
  {Li}}{{Lovell} et~al.}{2021}]{cite_lovell2021}
{Lovell} C.~C.,  {Geach} J.~E.,  {Dav{\'e}} R.,  {Narayanan} D.,   {Li} Q.,
  2021, \mn@doi [\mnras] {10.1093/mnras/staa4043}, \href
  {https://ui.adsabs.harvard.edu/abs/2021MNRAS.502..772L} {502, 772}

\bibitem[\protect\citeauthoryear{Misra}{Misra}{2019}]{cite_mish}
Misra D.,  2019, CoRR, abs/1908.08681

\bibitem[\protect\citeauthoryear{{Mustafa}, {Bard}, {Bhimji}, {Luki{\'c}},
  {Al-Rfou}  \& {Kratochvil}}{{Mustafa} et~al.}{2019}]{cite_mustafa2017}
{Mustafa} M.,  {Bard} D.,  {Bhimji} W.,  {Luki{\'c}} Z.,  {Al-Rfou} R.,
  {Kratochvil} J.~M.,  2019, \mn@doi [Computational Astrophysics and Cosmology]
  {10.1186/s40668-019-0029-9}, \href
  {https://ui.adsabs.harvard.edu/abs/2019ComAC...6....1M} {6, 1}

\bibitem[\protect\citeauthoryear{Nichol \& Dhariwal}{Nichol \&
  Dhariwal}{2021}]{cite_nichol2021}
Nichol A.,  Dhariwal P.,  2021, CoRR, abs/2102.09672

\bibitem[\protect\citeauthoryear{Paszke et~al.,}{Paszke
  et~al.}{2019}]{cite_pytorch}
Paszke A.,  et~al., 2019, in Wallach H.,  Larochelle H.,  Beygelzimer A.,
  d\textquotesingle Alch\'{e}-Buc F.,  Fox E.,   Garnett R.,  eds, , Advances
  in Neural Information Processing Systems 32.
Curran Associates, Inc., pp 8024--8035, \url
  {http://papers.neurips.cc/paper/9015-pytorch-an-imperative-style-high-performance-deep-learning-library.pdf}

\bibitem[\protect\citeauthoryear{Preechakul, Chatthee, Wizadwongsa  \&
  Suwajanakorn}{Preechakul et~al.}{2021}]{cite_preechakul2021}
Preechakul K.,  Chatthee N.,  Wizadwongsa S.,   Suwajanakorn S.,  2021, CoRR,
  abs/2111.15640

\bibitem[\protect\citeauthoryear{{Ravanbakhsh}, {Lanusse}, {Mandelbaum},
  {Schneider}  \& {Poczos}}{{Ravanbakhsh} et~al.}{2016}]{cite_ravanbakhsh2016}
{Ravanbakhsh} S.,  {Lanusse} F.,  {Mandelbaum} R.,  {Schneider} J.,   {Poczos}
  B.,  2016, arXiv e-prints, \href
  {https://ui.adsabs.harvard.edu/abs/2016arXiv160905796R} {p. arXiv:1609.05796}

\bibitem[\protect\citeauthoryear{{Reiman} \& {G{\"o}hre}}{{Reiman} \&
  {G{\"o}hre}}{2019}]{cite_reiman2019}
{Reiman} D.~M.,  {G{\"o}hre} B.~E.,  2019, \mn@doi [\mnras]
  {10.1093/mnras/stz575}, \href
  {https://ui.adsabs.harvard.edu/abs/2019MNRAS.485.2617R} {485, 2617}

\bibitem[\protect\citeauthoryear{{Remy}, {Lanusse}, {Ramzi}, {Liu}, {Jeffrey}
  \& {Starck}}{{Remy} et~al.}{2020}]{cite_remy2020}
{Remy} B.,  {Lanusse} F.,  {Ramzi} Z.,  {Liu} J.,  {Jeffrey} N.,   {Starck}
  J.-L.,  2020, arXiv e-prints, \href
  {https://ui.adsabs.harvard.edu/abs/2020arXiv201108271R} {p. arXiv:2011.08271}

\bibitem[\protect\citeauthoryear{Rezende \& Mohamed}{Rezende \&
  Mohamed}{2015}]{cite_rezende2015}
Rezende D.~J.,  Mohamed S.,  2015, in Proceedings of the 32nd International
  Conference on International Conference on Machine Learning - Volume 37.
  ICML'15.
JMLR.org, p. 1530–1538

\bibitem[\protect\citeauthoryear{Ronneberger, Fischer  \& Brox}{Ronneberger
  et~al.}{2015}]{cite_ronneberger2015}
Ronneberger O.,  Fischer P.,   Brox T.,  2015, in Navab N.,  Hornegger J.,  III
  W. M.~W.,   Frangi A.~F.,  eds,  Lecture Notes in Computer Science Vol. 9351,
  Medical Image Computing and Computer-Assisted Intervention - {MICCAI} 2015 -
  18th International Conference Munich, Germany, October 5 - 9, 2015,
  Proceedings, Part {III}. Springer, pp 234--241,
  \mn@doi{10.1007/978-3-319-24574-4\_28}, \url
  {https://doi.org/10.1007/978-3-319-24574-4\_28}

\bibitem[\protect\citeauthoryear{Saharia, Chan, Chang, Lee, Ho, Salimans, Fleet
   \& Norouzi}{Saharia et~al.}{2021}]{cite_saharia2021}
Saharia C.,  Chan W.,  Chang H.,  Lee C.~A.,  Ho J.,  Salimans T.,  Fleet
  D.~J.,   Norouzi M.,  2021, CoRR, abs/2111.05826

\bibitem[\protect\citeauthoryear{Salimans, Karpathy, Chen  \& Kingma}{Salimans
  et~al.}{2017}]{cite_salimans2017}
Salimans T.,  Karpathy A.,  Chen X.,   Kingma D.~P.,  2017, in 5th
  International Conference on Learning Representations, {ICLR} 2017, Toulon,
  France, April 24-26, 2017, Conference Track Proceedings. OpenReview.net, \url
  {https://openreview.net/forum?id=BJrFC6ceg}

\bibitem[\protect\citeauthoryear{Sasaki, Willcocks  \& Breckon}{Sasaki
  et~al.}{2021}]{cite_sasaki2021}
Sasaki H.,  Willcocks C.~G.,   Breckon T.~P.,  2021, CoRR, abs/2104.05358

\bibitem[\protect\citeauthoryear{{Schawinski}, {Zhang}, {Zhang}, {Fowler}  \&
  {Santhanam}}{{Schawinski} et~al.}{2017}]{cite_schawinski2017}
{Schawinski} K.,  {Zhang} C.,  {Zhang} H.,  {Fowler} L.,   {Santhanam} G.~K.,
  2017, \mn@doi [\mnras] {10.1093/mnrasl/slx008}, \href
  {https://ui.adsabs.harvard.edu/abs/2017MNRAS.467L.110S} {467, L110}

\bibitem[\protect\citeauthoryear{{Schaye} et~al.,}{{Schaye}
  et~al.}{2015}]{cite_eagle}
{Schaye} J.,  et~al., 2015, \mn@doi [\mnras] {10.1093/mnras/stu2058}, \href
  {https://ui.adsabs.harvard.edu/abs/2015MNRAS.446..521S} {446, 521}

\bibitem[\protect\citeauthoryear{Seitzer}{Seitzer}{2020}]{cite_seitzer2020}
Seitzer M.,  2020, {pytorch-fid: FID Score for PyTorch},
  \url{https://github.com/mseitzer/pytorch-fid}

\bibitem[\protect\citeauthoryear{{Smith} \& {Geach}}{{Smith} \&
  {Geach}}{2019}]{cite_smith2019}
{Smith} M.~J.,  {Geach} J.~E.,  2019, \mn@doi [\mnras] {10.1093/mnras/stz2886},
  \href {https://ui.adsabs.harvard.edu/abs/2019MNRAS.490.4985S} {490, 4985}

\bibitem[\protect\citeauthoryear{Sohl-Dickstein, Weiss, Maheswaranathan  \&
  Ganguli}{Sohl-Dickstein et~al.}{2015}]{cite_sohldickstein2015}
Sohl-Dickstein J.,  Weiss E.,  Maheswaranathan N.,   Ganguli S.,  2015, in Bach
  F.,  Blei D.,  eds,  Proceedings of Machine Learning Research Vol. 37,
  Proceedings of the 32nd International Conference on Machine Learning. PMLR,
  Lille, France, pp 2256--2265, \url
  {http://proceedings.mlr.press/v37/sohl-dickstein15.html}

\bibitem[\protect\citeauthoryear{{Somerville} \& {Primack}}{{Somerville} \&
  {Primack}}{1999}]{cite_somerville1999}
{Somerville} R.~S.,  {Primack} J.~R.,  1999, \mn@doi [\mnras]
  {10.1046/j.1365-8711.1999.03032.x}, \href
  {https://ui.adsabs.harvard.edu/abs/1999MNRAS.310.1087S} {310, 1087}

\bibitem[\protect\citeauthoryear{Song \& Ermon}{Song \&
  Ermon}{2019}]{cite_song2019}
Song Y.,  Ermon S.,  2019, in Wallach H.,  Larochelle H.,  Beygelzimer A.,
  d\textquotesingle Alch\'{e}-Buc F.,  Fox E.,   Garnett R.,  eds, ~ Vol. 32,
  Advances in Neural Information Processing Systems. Curran Associates, Inc.,
  \url
  {https://proceedings.neurips.cc/paper/2019/file/3001ef257407d5a371a96dcd947c7d93-Paper.pdf}

\bibitem[\protect\citeauthoryear{Song \& Ermon}{Song \&
  Ermon}{2020}]{cite_song2020}
Song Y.,  Ermon S.,  2020, in Larochelle H.,  Ranzato M.,  Hadsell R.,  Balcan
  M.~F.,   Lin H.,  eds, ~ Vol. 33, Advances in Neural Information Processing
  Systems. Curran Associates, Inc., pp 12438--12448, \url
  {https://proceedings.neurips.cc/paper/2020/file/92c3b916311a5517d9290576e3ea37ad-Paper.pdf}

\bibitem[\protect\citeauthoryear{Song, Sohl-Dickstein, Kingma, Kumar, Ermon  \&
  Poole}{Song et~al.}{2021}]{cite_song2021}
Song Y.,  Sohl-Dickstein J.,  Kingma D.~P.,  Kumar A.,  Ermon S.,   Poole B.,
  2021, in International Conference on Learning Representations. \url
  {https://openreview.net/forum?id=PxTIG12RRHS}

\bibitem[\protect\citeauthoryear{{Spindler}, {Geach}  \& {Smith}}{{Spindler}
  et~al.}{2021}]{cite_spindler2021}
{Spindler} A.,  {Geach} J.~E.,   {Smith} M.~J.,  2021, \mn@doi [\mnras]
  {10.1093/mnras/staa3670}, \href
  {https://ui.adsabs.harvard.edu/abs/2021MNRAS.502..985S} {502, 985}

\bibitem[\protect\citeauthoryear{Srivastava, Greff  \& Schmidhuber}{Srivastava
  et~al.}{2015}]{cite_srivastava2015}
Srivastava R.~K.,  Greff K.,   Schmidhuber J.,  2015, CoRR, abs/1505.00387

\bibitem[\protect\citeauthoryear{{Stark} et~al.,}{{Stark}
  et~al.}{2018}]{cite_stark2018}
{Stark} D.,  et~al., 2018, \mn@doi [\mnras] {10.1093/mnras/sty764}, \href
  {https://ui.adsabs.harvard.edu/abs/2018MNRAS.477.2513S} {477, 2513}

\bibitem[\protect\citeauthoryear{{Stone} \& {Courteau}}{{Stone} \&
  {Courteau}}{2019}]{cite_stone2019}
{Stone} C.,  {Courteau} S.,  2019, \mn@doi [\apj] {10.3847/1538-4357/ab3126},
  \href {https://ui.adsabs.harvard.edu/abs/2019ApJ...882....6S} {882, 6}

\bibitem[\protect\citeauthoryear{{Stone}, {Courteau}  \& {Arora}}{{Stone}
  et~al.}{2021}]{cite_stone2021a}
{Stone} C.,  {Courteau} S.,   {Arora} N.,  2021, \mn@doi [\apj]
  {10.3847/1538-4357/abebe4}, \href
  {https://ui.adsabs.harvard.edu/abs/2021ApJ...912...41S} {912, 41}

\bibitem[\protect\citeauthoryear{Strubell, Ganesh  \& McCallum}{Strubell
  et~al.}{2019}]{cite_strubell2019}
Strubell E.,  Ganesh A.,   McCallum A.,  2019, in Proceedings of ACL 57.
  Association for Computational Linguistics, Florence, Italy, p.~3645,
  \mn@doi{10.18653/v1/P19-1355}

\bibitem[\protect\citeauthoryear{Szegedy, Vanhoucke, Ioffe, Shlens  \&
  Wojna}{Szegedy et~al.}{2016}]{cite_szegedy2016}
Szegedy C.,  Vanhoucke V.,  Ioffe S.,  Shlens J.,   Wojna Z.,  2016, in
  Proceedings of the IEEE Conference on Computer Vision and Pattern Recognition
  (CVPR).

\bibitem[\protect\citeauthoryear{{Tamosiunas}, {Winther}, {Koyama}, {Bacon},
  {Nichol}  \& {Mawdsley}}{{Tamosiunas} et~al.}{2020}]{cite_tamosiunas2020}
{Tamosiunas} A.,  {Winther} H.~A.,  {Koyama} K.,  {Bacon} D.~J.,  {Nichol}
  R.~C.,   {Mawdsley} B.,  2020, arXiv e-prints, \href
  {https://ui.adsabs.harvard.edu/abs/2020arXiv200410223T} {p. arXiv:2004.10223}

\bibitem[\protect\citeauthoryear{{Trayford} et~al.,}{{Trayford}
  et~al.}{2017}]{cite_trayford2017}
{Trayford} J.~W.,  et~al., 2017, \mn@doi [\mnras] {10.1093/mnras/stx1051},
  \href {https://ui.adsabs.harvard.edu/abs/2017MNRAS.470..771T} {470, 771}

\bibitem[\protect\citeauthoryear{Vincent}{Vincent}{2011}]{cite_vincent2011}
Vincent P.,  2011, \mn@doi [Neural Computation] {10.1162/NECO_a_00142}, 23,
  1661–1674

\bibitem[\protect\citeauthoryear{{Vogelsberger} et~al.,}{{Vogelsberger}
  et~al.}{2014}]{cite_vogelsberger2014}
{Vogelsberger} M.,  et~al., 2014, \mn@doi [\mnras] {10.1093/mnras/stu1536},
  \href {https://ui.adsabs.harvard.edu/abs/2014MNRAS.444.1518V} {444, 1518}

\bibitem[\protect\citeauthoryear{{Wilman}, {Zibetti}  \&
  {Budav{\'a}ri}}{{Wilman} et~al.}{2010}]{cite_wilman2010}
{Wilman} D.~J.,  {Zibetti} S.,   {Budav{\'a}ri} T.,  2010, \mn@doi [\mnras]
  {10.1111/j.1365-2966.2010.16845.x}, \href
  {https://ui.adsabs.harvard.edu/abs/2010MNRAS.406.1701W} {406, 1701}

\bibitem[\protect\citeauthoryear{{York} et~al.,}{{York}
  et~al.}{2000}]{cite_sdss}
{York} D.~G.,  et~al., 2000, \mn@doi [\aj] {10.1086/301513}, \href
  {https://ui.adsabs.harvard.edu/abs/2000AJ....120.1579Y} {120, 1579}

\makeatother
\end{thebibliography}
\bsp	

\label{lastpage}
\end{document}